\documentclass[runningheads]{llncs}
\usepackage[T1]{fontenc}

\usepackage{amssymb}
\usepackage{subcaption}
\usepackage{comment}
\usepackage{float}
\usepackage{siunitx}
\PassOptionsToPackage{hyphens}{url}\usepackage{hyperref}
\usepackage[french]{babel}
\usepackage{cleveref}
\usepackage[utf8]{inputenc}
\usepackage[right]{lineno}
\usepackage{csquotes}
\usepackage{booktabs}
\usepackage{longtable}
\usepackage{adjustbox}
\usepackage{array}
\usepackage{url}
\usepackage{authblk}
\usepackage{xcolor} % Load the xcolor package for color options
\usepackage{caption}
\DeclareCaptionLabelFormat{default}{#1~#2}
\title{Prédiction du risque d'échec dans un MOOC : une approche basée sur l'analyse des séries temporelles multivariées}

\author{Anass El Ayady \inst{1,2} \and Maxime Devanne \inst{2} \and Germain Forestier \inst{2} \and Nour El Mawas \inst{1} }

\institute{
CREM, Université de Lorraine, Metz, France \\
\email{\{anass.el-ayady, nour.el-mawas\}@univ-lorraine.fr}
\and
IRIMAS, Université de Haute-Alsace, Mulhouse, France \\
\email{\{maxime.devanne, germain.forestier\}@uha.fr}
}

\titlerunning{Prédiction du risque d’échec dans les MOOCs}
\authorrunning{Environnements Informatiques pour l’Apprentissage Humain 2025}

\begin{document}
\maketitle

\begin{abstract}
Les MOOCs offrent un accès libre et gratuit à un large public, mais les taux de complétion restent faibles, souvent en raison du manque de personnalisation. Pour y remédier, il est essentiel de prédire les performances des apprenants afin de leur fournir une rétroaction adaptée. Les traces comportementales (clics, événements) peuvent être analysées sous forme de séries temporelles pour anticiper leurs résultats. Ce travail compare des méthodes de classification de séries temporelles multivariées pour identifier les apprenants à risque d’échec à différents moments du cours (après 5, 10 semaines, etc.). L’évaluation expérimentale, menée sur la base \textit{Open University Learning Analytics Dataset} (OULAD), porte sur trois cours : deux en \textbf{STEM} et un en \textbf{SHS}. Les résultats préliminaires montrent que les approches évaluées sont prometteuses pour prédire l’échec dans les MOOCs. Les analyses suggèrent aussi que la précision dépend du volume d’interactions, soulignant l’importance de données comportementales riches et variées.
%add 6 keywords
\end{abstract}
%\linenumbers
\renewcommand{\abstractname}{Abstract}
\begin{abstract}
MOOCs offer free and open access to a wide audience, but completion rates remain low, often due to a lack of personalized content. To address this issue, it is essential to predict learner performance in order to provide tailored feedback. Behavioral traces—such as clicks and events—can be analyzed as time series to anticipate learners' outcomes. This work compares multivariate time series classification methods to identify at-risk learners at different stages of the course (after 5, 10 weeks, etc.). The experimental evaluation, conducted on the Open University Learning Analytics Dataset (OULAD), focuses on three courses: two in STEM and one in SHS. Preliminary results show that the evaluated approaches are promising for predicting learner failure in MOOCs. The analysis also suggests that prediction accuracy is influenced by the amount of recorded interactions, highlighting the importance of rich and diverse behavioral data.

\end{abstract}

\section{Introduction}
\label{sec:introduction}

Avec l'essor des technologies numériques, les MOOCs se sont imposés comme un outil clé de démocratisation de l’éducation. Ils permettent à des millions d’apprenants, de tous horizons, d’accéder à des contenus éducatifs variés. Pourtant, leur potentiel est souvent freiné par des taux d’abandon élevés qui varient généralement entre 0,7\% et 52,1\%, avec une médiane de 12,6\% \cite{jordan2015massive}. Cela souligne un défi persistant dans l’adaptation des cours aux besoins spécifiques des apprenants.

En parallèle, l’évolution des outils d’intelligence artificielle et des méthodes d’apprentissage automatique offre de nouvelles opportunités pour résoudre ces problématiques. Les approches modernes basées sur l’analyse des données permettent non seulement d’automatiser l’identification des facteurs de risque, mais également d’adopter une approche proactive pour engager les apprenants. En mobilisant des algorithmes sophistiqués, il devient possible de proposer des contenus adaptés et des stratégies pédagogiques personnalisées, maximisant ainsi l’expérience éducative.

L'intégration de méthodes d'analyse des données provenant des interactions des apprenants avec la plateforme pourrait constituer une réponse prometteuse à ce défi. Ces données, parfois négligées, représentent une véritable source d’informations sur les comportements et les trajectoires des apprenants au fil du temps. En particulier, les clics de souris peuvent être exploités pour modéliser des séries temporelles décrivant leurs interactions avec différentes activités pédagogiques. Ces séries permettent de suivre les rythmes d’apprentissage, d’identifier les tendances et de prédire les risques d’échec avant qu’ils ne deviennent critiques.

Ce travail se distingue par l’utilisation des séries temporelles multivariées, rarement exploitées dans ce contexte, et l’introduction du modèle \textit{Fully Convolutional Networks} (FCN) pour la prédiction du risque d’échec dans les MOOCs. Par ailleurs, il propose une analyse différenciée selon le type de cours (STEM vs SHS), un aspect encore peu exploré dans la littérature. L’objectif est d’identifier les étudiants à risque dès les premières semaines afin de favoriser des interventions pédagogiques ciblées.

\section{Cadre théorique}
\label{sec:method}

Les recherches sur la prédiction des performances des apprenants dans les environnements d’apprentissage en ligne s’appuient de plus en plus sur des approches issues de l’intelligence artificielle, de l’apprentissage automatique et de l’apprentissage profond. Le succès récent des modèles d’apprentissage profond motive leur adoption croissante pour exploiter les données disponibles, notamment dans des bases comme OULAD \cite{kuzilek2017open}, reconnue pour sa richesse et son accessibilité.

Les modèles récurrents, tels que ceux utilisant des LSTM, se sont montrés efficaces pour capturer les dépendances temporelles dans les données issues d'OULAD. Par exemple, le modèle DOPP \cite{karimi2020deep} a démontré son utilité pour identifier les apprenants à risque dès les premières semaines, ce qui le rend pertinent pour une intervention précoce. Une approche hybride combinant des couches denses et des LSTM a également été proposée pour la prédiction multiclasses, illustrant la capacité de ces modèles à traiter des tâches complexes. De plus, les modèles GRU et LSTM étudiés dans \cite{he2020online} et \cite{aljohani2019predicting} confirment leur pertinence pour analyser les séquences temporelles et prédire les risques d'échec des apprenants, en particulier dans des contextes nécessitant des prédictions précoces et fiables.

D'autres méthodologies exploitent des relations complexes entre apprenants et cours. Une approche basée sur des graphes neuronaux, combinée à des modèles récurrents, représente ces relations sous forme de graphes de connaissances \cite{karimi2020online}. En utilisant des données de clics et des attributs relationnels temporels, cette méthode démontre son efficacité pour effectuer des prédictions précoces dans un contexte multi-relationnel. Ces travaux mettent en avant l'intérêt d'explorer des représentations graphiques pour améliorer la compréhension et la prédiction dans les systèmes éducatifs.

Les modèles basés sur des Transformers \cite{kusumawardani2023transformer} se distinguent par leur capacité à capturer des dépendances séquentielles grâce à des mécanismes d’attention. Ces approches permettent de traiter des données normalisées et organisées de manière séquentielle pour des prédictions multiclasses. En exploitant des données agrégées à différentes granularités temporelles, les Transformers démontrent leur efficacité pour les tâches de classification séquentielle, surpassant souvent les modèles traditionnels comme les LSTM.

Des méthodes plus classiques exploitent les forêts de séries temporelles pour classer les apprenants en fonction de leurs comportements interactifs \cite{haiyang2018time}. De manière similaire, des modèles d’apprentissage automatique plus simples, comme les forêts aléatoires, permettent des prédictions précoces en analysant une partie limitée des données disponibles. Ces travaux illustrent la robustesse des algorithmes classiques lorsqu’ils sont appliqués à des données bien structurées.

Comparativement, les modèles récurrents (LSTM, GRU) excellent dans l’analyse des données temporelles continues et les prédictions à différents moments d’un cours grâce à leur capacité à modéliser les séquences. Les approches par graphes et Transformers s’avèrent efficaces pour capturer des relations complexes entre apprenants et contenus. Les modèles classiques, tels que les forêts aléatoires, bien que robustes et simples, sont moins adaptés aux prédictions séquentielles ou aux contextes relationnels complexes.

Cependant, ces avancées ne prennent pas en compte l'influence des différences disciplinaires, notamment entre les cours en SHS et en STEM. Ces distinctions, pourtant essentielles, peuvent affecter la densité des données et les performances des modèles prédictifs. Notre étude comble cette lacune en analysant ces deux contextes distincts et en évaluant les approches pour chaque discipline.

Dans notre étude, nous avons sélectionné les modèles LSTM pour leur performance reconnue en analyse de séries temporelles, comme indiqué dans la littérature. Le modèle DOPP, basé sur LSTM, est réputé pour capturer efficacement les dépendances temporelles et fournir des prédictions précoces à partir des données de clics.

En complément, nous avons exploré le FCN, qui se distingue par sa capacité à extraire des motifs locaux et globaux dans les séries temporelles grâce à ses couches convolutives. Bien qu’ils soient moins présents dans les travaux de l'état de l'art sur les MOOCs, ils ont démontré leur efficacité dans la classification des séries temporelles dans d'autres domaines \cite{ismail2019deep}, ce qui justifie leur inclusion dans notre étude.

Pour compléter cette sélection, nous avons intégré des modèles classiques comme les k-plus-proches-voisins (KNN) et les perceptrons multicouches (MLP). Bien qu’ils ne soient pas couramment étudiés dans le domaine de la prédiction des performances dans les MOOCs, ces modèles offrent une base de comparaison pour évaluer la pertinence et la valeur ajoutée des approches avancées. L’objectif de cette étude comparative est d’évaluer différentes méthodes d’analyse des séries temporelles, dans des conditions cohérentes, afin d’identifier celles qui semblent les plus prometteuses pour prédire le risque d’échec dans les contextes ciblés. En particulier, les modèles avancés comme le FCN et le DOPP exploitent les dépendances temporelles et les séries multivariées, tandis que les KNN et les MLP permettent d’évaluer la performance de solutions plus simples. Ce choix équilibré entre approches classiques et avancées permet d’obtenir une vue d’ensemble complète et rigoureuse des méthodes disponibles pour ce contexte spécifique.

\section{Méthodologie}

Cette section détaille les approches méthodologiques adoptées pour analyser les séries temporelles des interactions des apprenants et prédire leur niveau de réussite. Nous décrivons d’abord le jeu de données utilisé, puis les étapes de préparation des données, les modèles évalués, ainsi que les méthodologies d’évaluation utilisées pour comparer leurs performances.

\subsection{Description du jeu de données} 
Cette étude s’appuie sur le jeu de données public OULAD, qui regroupe les parcours de 32 593 étudiants répartis sur 22 cours en ligne, avec plus de 10 millions de clics enregistrés. Nous avons sélectionné trois cours fortement représentés (tableau \ref{tab:statis}) : (BBB, SHS), (DDD, STEM) et (FFF, STEM), afin d’explorer l’influence du type disciplinaire sur les performances prédictives.

\begin{table*}[h]
\centering
\caption{Statistiques des apprenants et des présentations pour chaque module}
\begin{tabular}{ccc}
\toprule
\textbf{Code de module} & \textbf{Nombre de présentations} & \textbf{Nombre d'apprenants} \\
\midrule
AAA & 2 & 748 \\
BBB & 4 & 7909 \\
CCC & 2 & 4434 \\
DDD & 4 & 6272 \\
EEE & 3 & 2934 \\
FFF & 4 & 7762 \\
GGG & 3 & 2534 \\
\midrule
\textbf{Total} & \textbf{22} & \textbf{32593} \\
\bottomrule
\end{tabular}
\label{tab:statis}
\end{table*}

Chaque cours est proposé deux fois par an (sessions de février "B" et d'octobre "J") (tableau \ref{tab:courses}). Les sessions 2013B et 2013J sont utilisées pour l'entraînement, et les sessions 2014B et 2014J pour le test, ce qui garantit une évaluation sur des cohortes indépendantes dans le temps. Certains apprenants inscrits ne participent pas au cours, ce qui explique la différence entre le nombre total d'apprenants indiqué dans le tableau~\ref{tab:statis} et la somme des apprenants par période dans le tableau~\ref{tab:courses}. 

\begin{table}[h]
    \centering
    \caption{Description des cours étudiés}
    \begin{tabular}{>{\centering\arraybackslash}p{2cm} >{\centering\arraybackslash}p{3cm} >{\centering\arraybackslash}p{2.5cm} >{\centering\arraybackslash}p{4cm}}
        \toprule
        \textbf{Code} & \textbf{Type} & \textbf{Période} & \textbf{Nombre d'apprenants} \\ \midrule
        BBB & SHS & 2013B & 1537 \\ 
        BBB & SHS & 2013J & 1870 \\ 
        BBB & SHS & 2014B & 1294 \\ 
        BBB & SHS & 2014J & 1921 \\ 
        DDD & STEM & 2013B & 1214 \\ 
        DDD & STEM & 2013J & 1768 \\ 
        DDD & STEM & 2014B & 1116 \\ 
        DDD & STEM & 2014J & 1647 \\ 
        FFF & STEM & 2013B & 1510 \\ 
        FFF & STEM & 2013J & 2098 \\ 
        FFF & STEM & 2014B & 1363 \\ 
        FFF & STEM & 2014J & 2121 \\ 
        \bottomrule
    \end{tabular}
    \label{tab:courses}
\end{table}

La distribution des résultats finaux est la suivante : \textit{Pass} (42,55\%), \textit{Withdrawn} (24,62\%), \textit{Fail} (23,94\%), \textit{Distinction} (8,89\%). Pour la classification binaire, \textit{Pass} et \textit{Distinction} sont regroupés en classe positive, tandis que \textit{Fail} et \textit{Withdrawn} forment la classe négative. Une baseline simple basée sur la classe majoritaire atteint un F1-score de 0,58 (binaire) et 0,48 (multiclasse), ce qui reste significativement inférieur aux performances obtenues par les modèles proposés.

\subsection{Préparation des données} 
Les interactions des étudiants avec les cours sont transformées en séries temporelles multivariées, où chaque étudiant est décrit par une séquence hebdomadaire de clics répartis par type d’activité (forum, quiz, ressource, etc.). Chaque activité constitue un \textit{canal}, soit une dimension distincte de la série.

La structure obtenue est $(n\_samples,\ n\_weeks,\ n\_activities)$, où $n\_samples$ est le nombre d’apprenants, $n\_weeks$ la durée d’observation (variant de 5 à 40 semaines) et $n\_activities$ le nombre d’activités distinctes. Pour garantir l’homogénéité des canaux, une normalisation Min-Max est appliquée individuellement sur chaque canal.

Les étiquettes de performance sont assignées selon deux cadres : \textit{binaire} (Pass/Fail) et \textit{multiclasse} (Distinction, Pass, Fail, Withdrawn), permettant d’évaluer les modèles à différents niveaux de granularité.

\begin{figure}[ht]
    \centering
    \includegraphics[height=0.35\textheight]{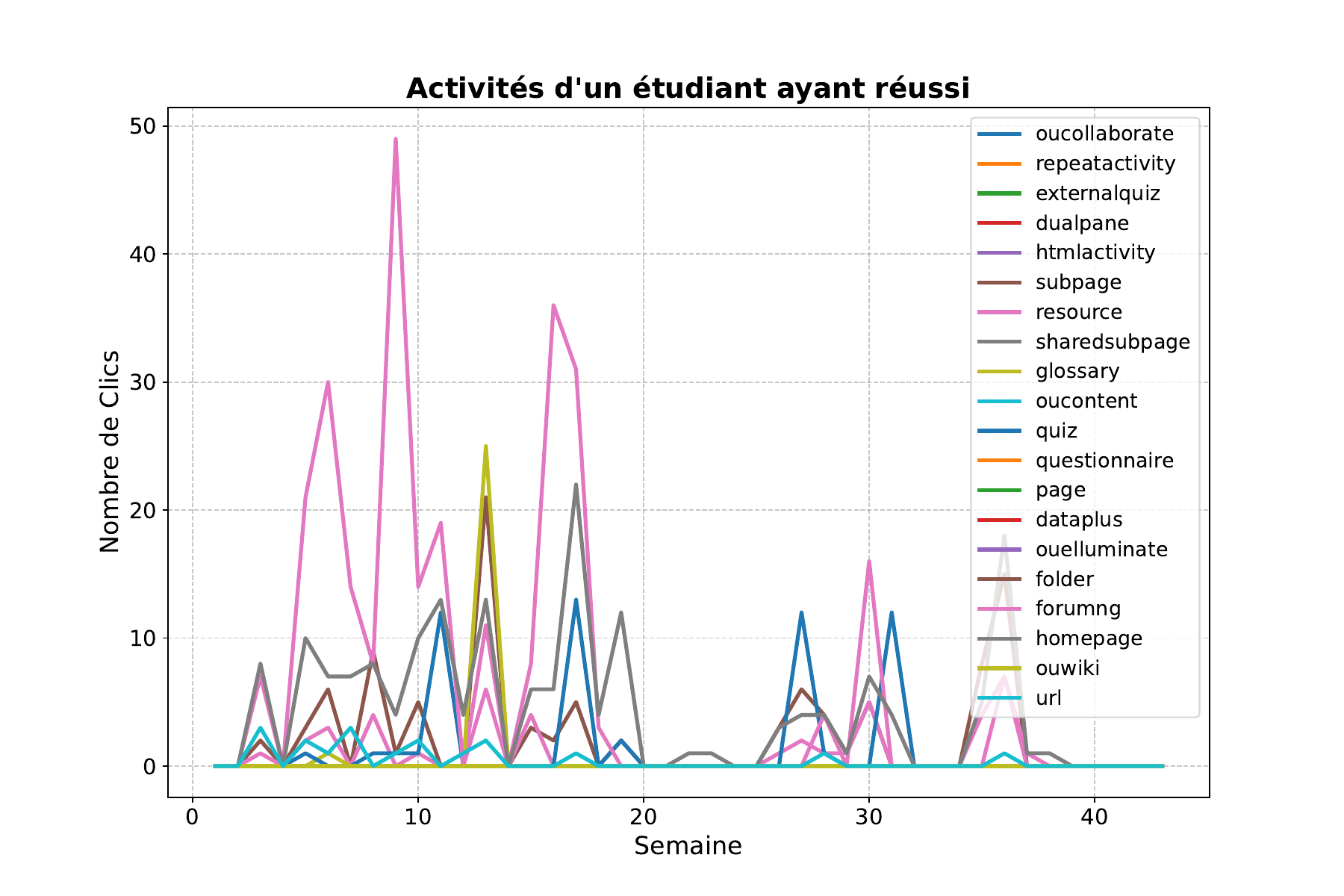}
    \caption{Représentation des interactions hebdomadaires d’un apprenant ayant réussi. L’activité reste régulière tout au long du cours, indiquant un engagement constant.}
    \label{fig:pass} 
\end{figure}

\begin{figure}[ht]
    \centering
    \includegraphics[height=0.35\textheight]{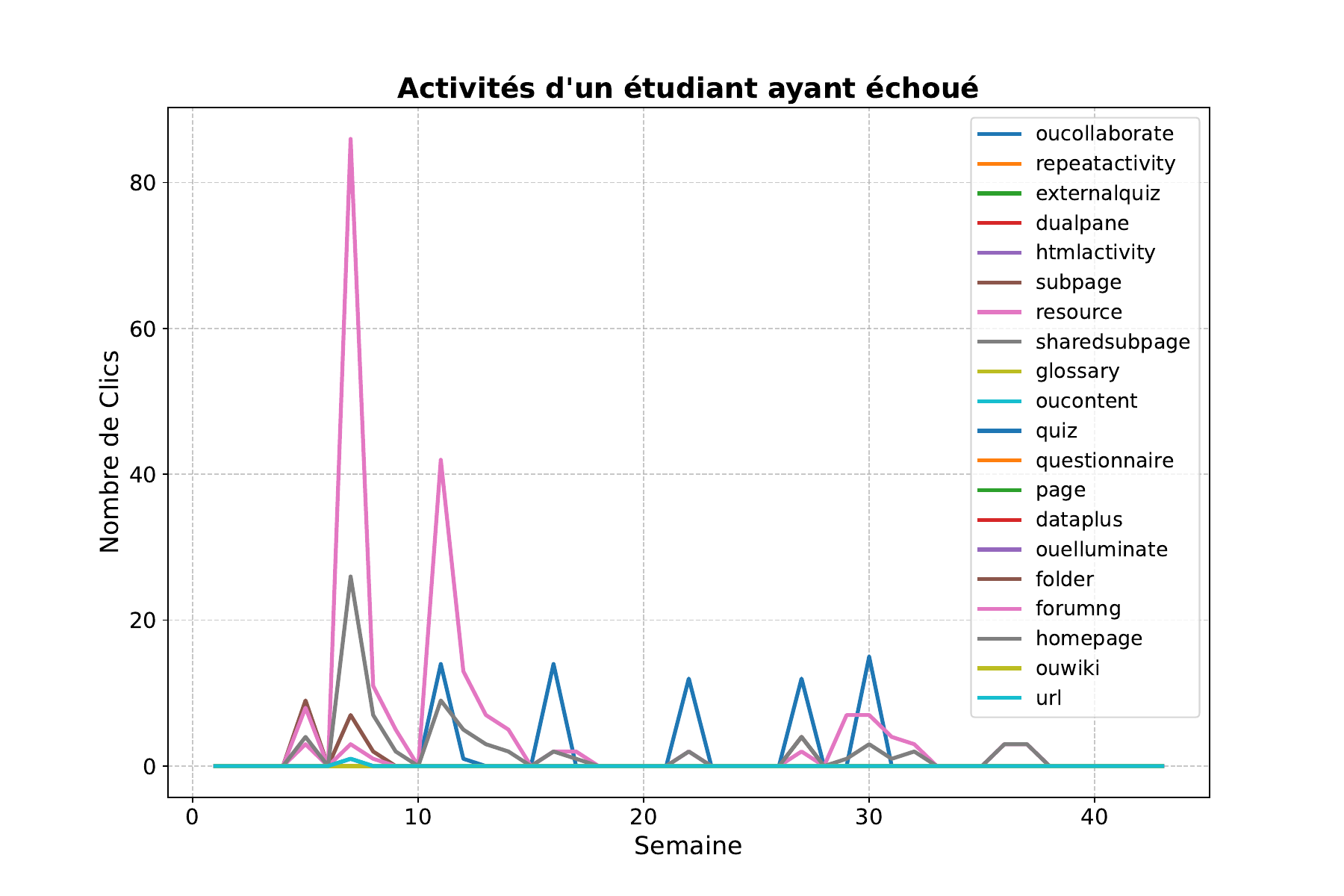}
    \caption{Représentation des interactions hebdomadaires d’un apprenant échoué. Une baisse significative de l’activité est observée à partir de la 5e semaine, signalant un désengagement progressif.}
    \label{fig:fail} 
\end{figure}

La Fig.~\ref{fig:pass} montre que les apprenants qui réussissent maintiennent une activité régulière tout au long du cours, tandis que la Fig.~\ref{fig:fail} illustre une baisse significative des interactions pour les apprenants en échec. Ces dynamiques confirment l’importance des traces d’interactions comme indicateurs précoces.

\subsection{Indicateur de performance}

L’évaluation repose sur le \textit{F1-score}, une métrique équilibrée combinant précision et rappel. Il est particulièrement adapté aux contextes de classification déséquilibrée, comme les MOOCs, où la simple précision globale (accuracy) peut être trompeuse.

Le F1-score est défini par :
\begin{equation}
F1 = \frac{2 \times \text{Précision} \times \text{Rappel}}{\text{Précision} + \text{Rappel}}
\end{equation}

avec :
\begin{equation}
\text{Précision} = \frac{TP}{TP + FP}, \quad \text{Rappel} = \frac{TP}{TP + FN}
\end{equation}

où $TP$, $FP$ et $FN$ désignent respectivement les vrais positifs, les faux positifs et les faux négatifs. Dans notre contexte, cela reflète la capacité du modèle à détecter correctement les apprenants à risque, tout en limitant les fausses alertes.

Les performances sont mesurées à différents moments du cours (après 5, 10, 15 semaines, etc.), afin d’évaluer la capacité des modèles à prédire tôt dans le parcours. Cette approche permet d’identifier les périodes critiques où les prédictions sont les plus fiables.

%\newpage
\section{Résultats et Discussion}
\begin{figure}[ht]
    \centering
    % Deux figures côte à côte
    \begin{subfigure}[b]{0.49\textwidth}
        \includegraphics[width=\textwidth]{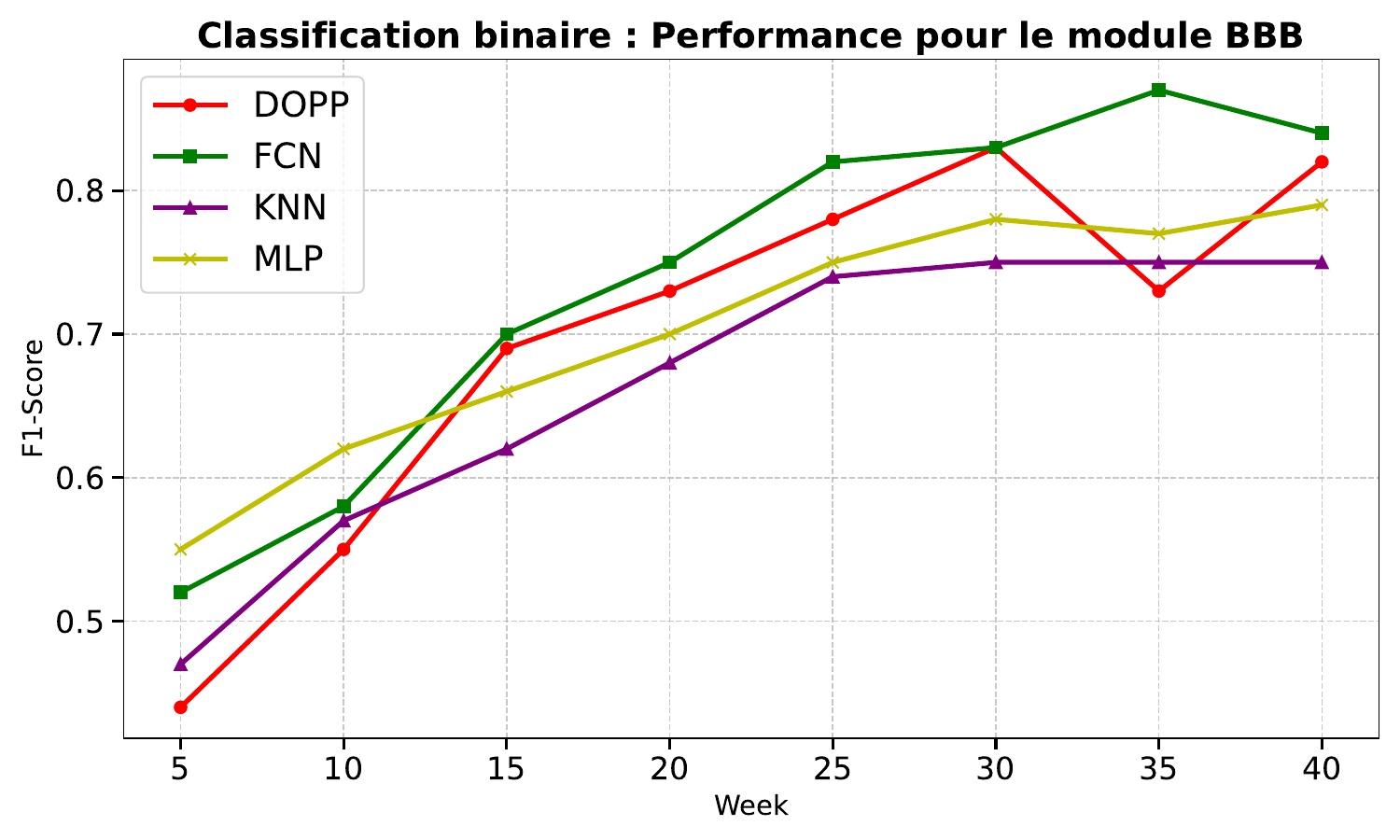}
        \caption{Le cours BBB}
        \label{fig:1}
    \end{subfigure}
    \hfill
    \begin{subfigure}[b]{0.49\textwidth}
        \includegraphics[width=\textwidth]{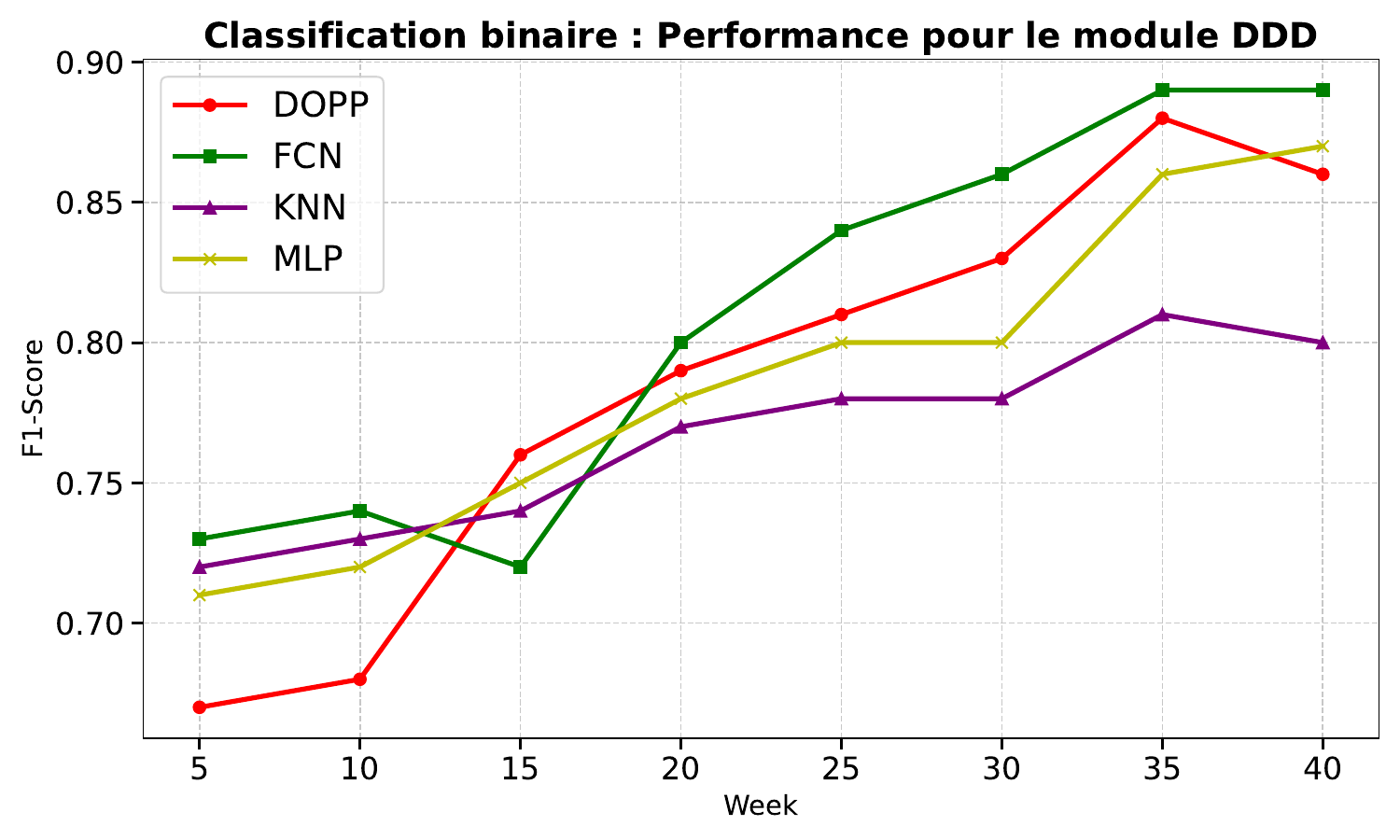}
        \caption{Le cours DDD}
        \label{fig:2}
    \end{subfigure}
    
    % Figure en dessous, centrée
    \begin{subfigure}[b]{0.49\textwidth}
        \centering
        \includegraphics[width=\textwidth]{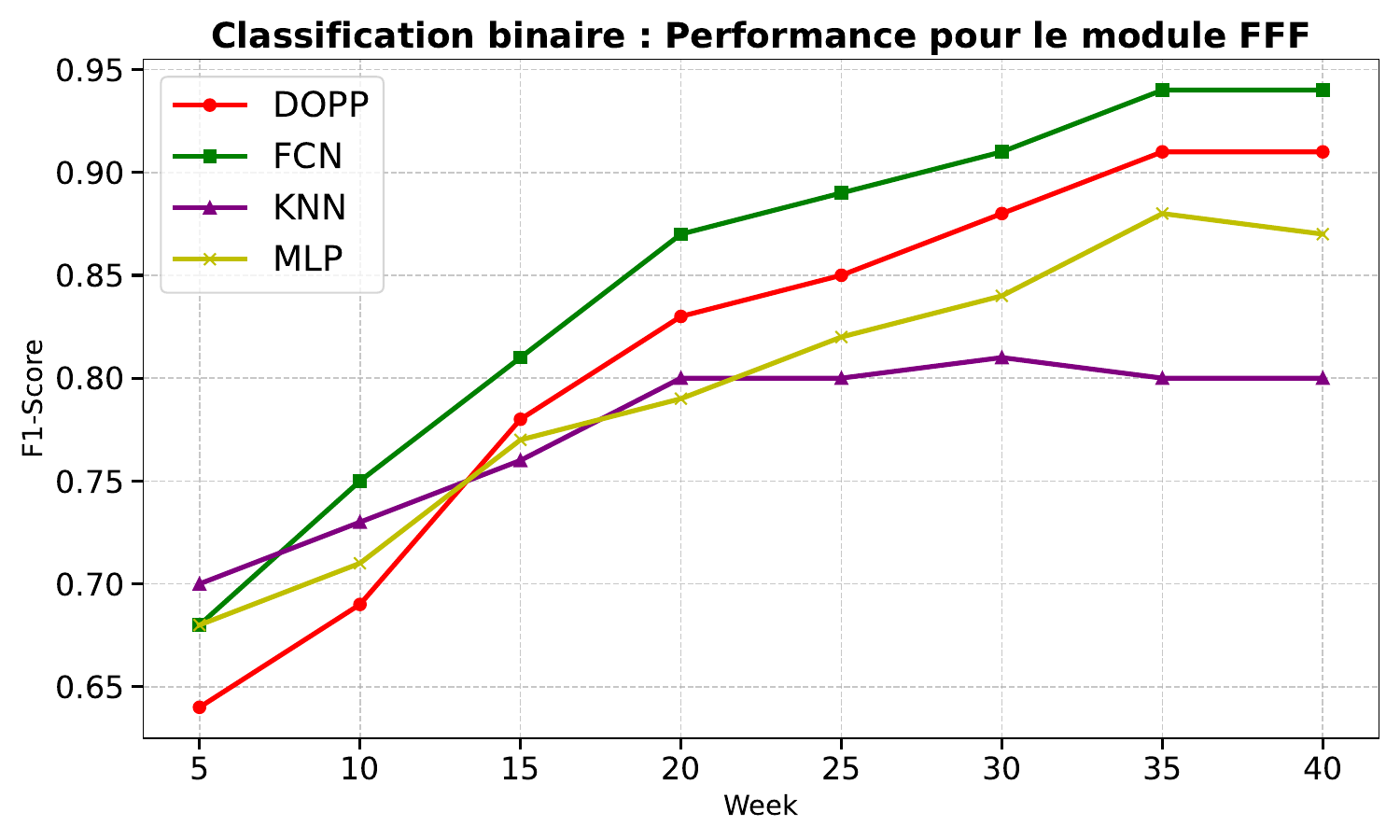}
        \caption{Le cours FFF}
        \label{fig:3}
    \end{subfigure}

    \caption{Évaluation de performances : classification binaire}
    \label{fig:all1}
\end{figure}

Les résultats obtenus pour la classification binaire et multiclasses révèlent des tendances similaires en termes de performances des modèles et de dépendance au type de cours et à la durée d’observation. Pour la \textbf{classification binaire} (Fig.~\ref{fig:all1}), les cours \textbf{STEM} (\textbf{DDD} et \textbf{FFF}) affichent des scores F1 significativement plus élevés que le cours \textbf{SHS} (\textbf{BBB}), atteignant jusqu’à \textbf{94\%} pour \textbf{FCN} après 40 semaines dans le cours FFF. Le cours \textbf{DDD}, bien qu’étant également un cours STEM, affiche des performances légèrement inférieures (86\%). Cette différence s’explique par un volume et une régularité d’interactions inférieurs à ceux observés dans FFF, comme le confirment les statistiques d’usage. Ce résultat souligne la capacité du \textbf{FCN} à identifier des motifs temporels complexes grâce à ses convolutions, qui permettent d’extraire des relations locales et globales dans les séries temporelles tout en préservant leur structure séquentielle.

Cette supériorité des modèles pour les cours STEM s’explique sûrement par le volume élevé d’interactions et une variabilité temporelle plus marquée. Cela offre aux modèles, notamment \textbf{FCN} et \textbf{DOPP}, une meilleure capacité d’apprentissage des relations complexes entre les activités et les résultats des apprenants. À l’inverse, le cours SHS souffre d’une stagnation des performances, en particulier pour les séries temporelles courtes, à cause d’une faible densité d’interactions : en configuration binaire, le F1-score reste limité à \textbf{58\%} après 5 semaines; en multiclasses, il débute à \textbf{32\%} après 5 semaines et plafonne à \textbf{54\%} après 40 semaines.

\begin{figure}[ht]
    \centering
    % Deux figures côte à côte
    \begin{subfigure}[b]{0.49\textwidth}
        \includegraphics[width=\textwidth]{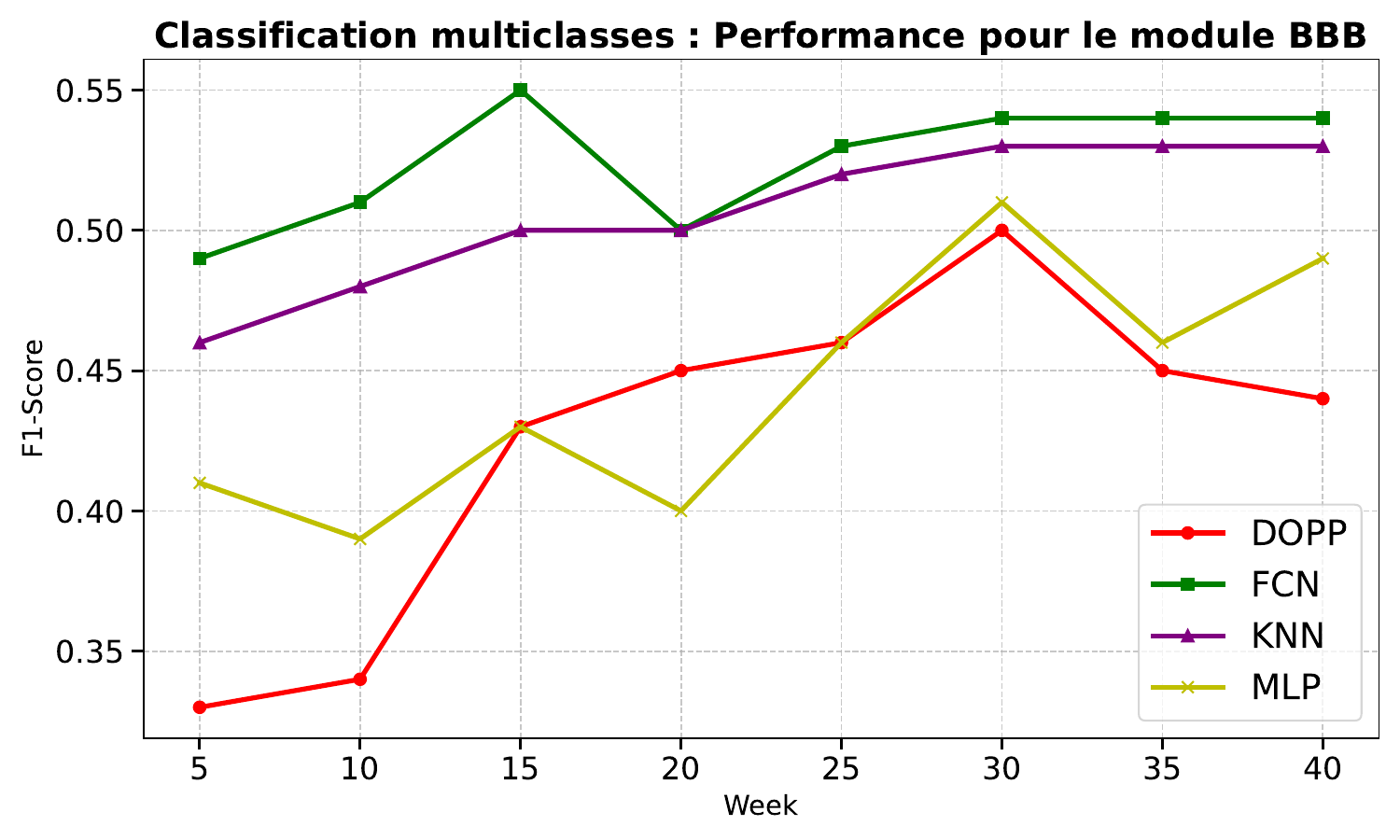}
        \caption{Le cours BBB}
        \label{fig:4}
    \end{subfigure}
    \hfill
    \begin{subfigure}[b]{0.49\textwidth}
        \includegraphics[width=\textwidth]{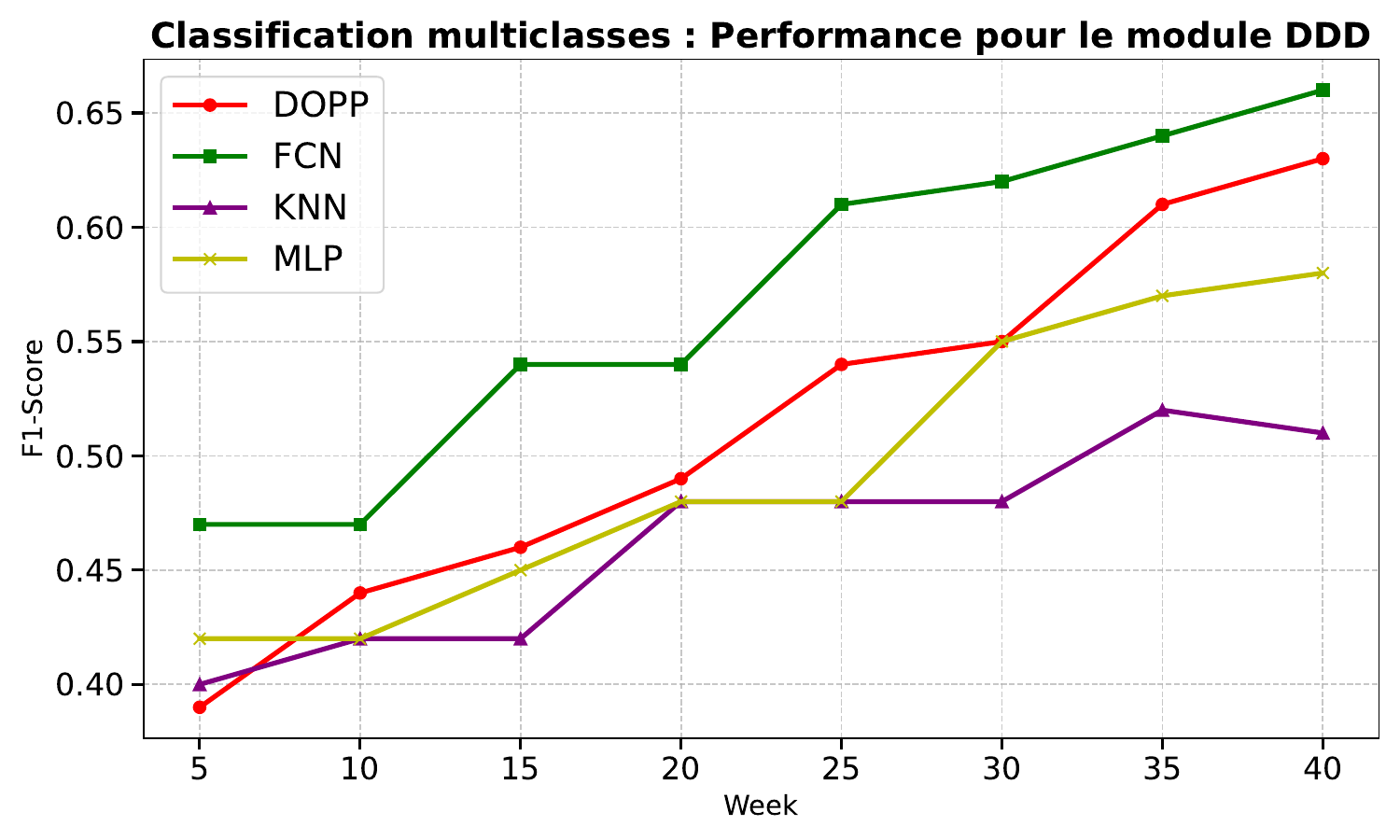}
        \caption{Le cours DDD}
        \label{fig:5}
    \end{subfigure}
    
    % Figure en dessous, centrée
   
    \begin{subfigure}[b]{0.49\textwidth}
        \centering
        \includegraphics[width=\textwidth]{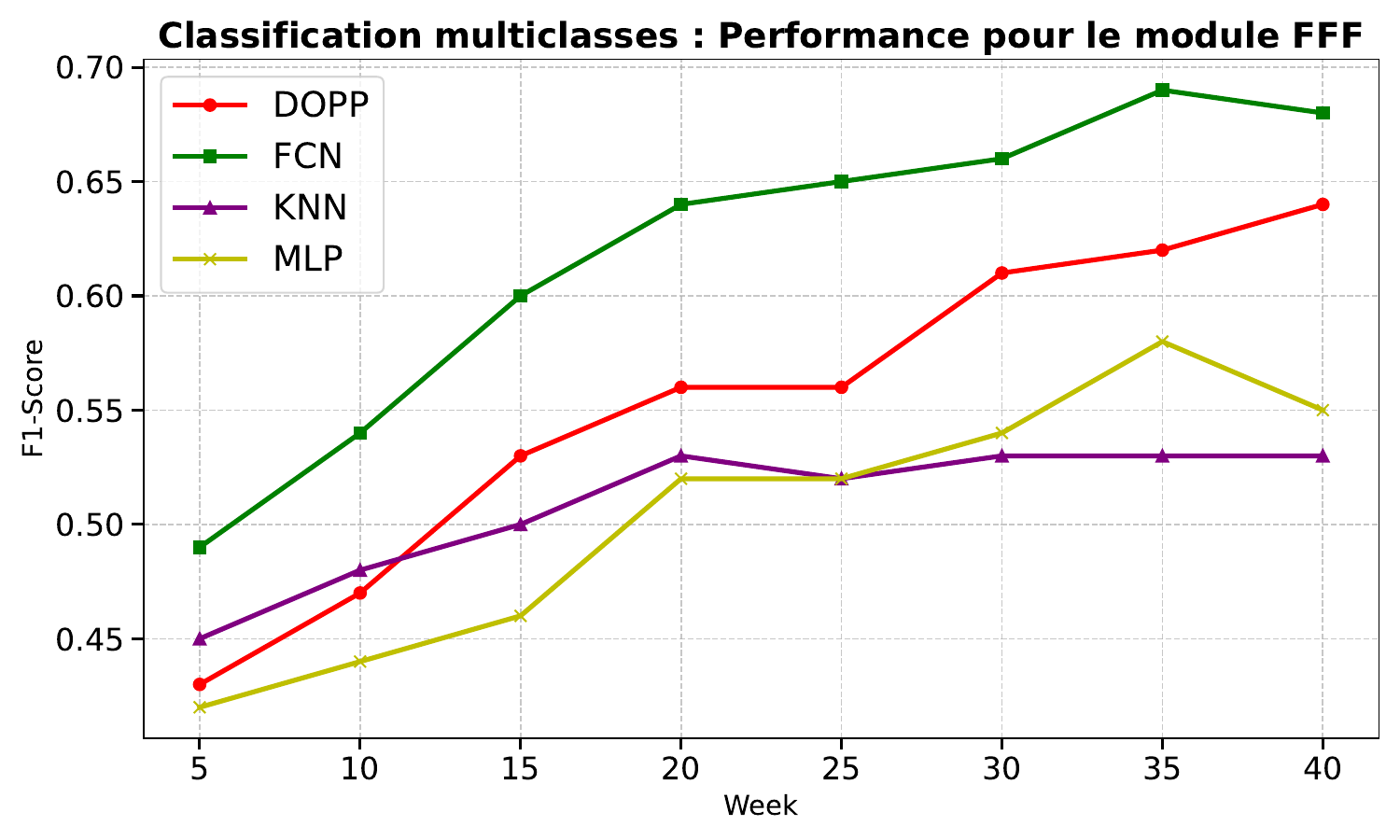}
        \caption{Le cours FFF}
        \label{fig:6}
    \end{subfigure}

    \caption{Évaluation de performances : classification multiclasse}
    \label{fig:all2}
\end{figure}

Pour la \textbf{classification multiclasses} (Fig.~\ref{fig:all2}), ces tendances persistent, mais les scores F1 sont globalement plus faibles, en particulier pour le cours \textbf{BBB}, où \textbf{FCN} plafonne autour de \textbf{55\%} après 40 semaines. Cela illustre la difficulté des modèles à discriminer plusieurs classes dans un environnement où les données interactives sont limitées. En revanche, pour les cours \textbf{DDD} et \textbf{FFF}, les modèles montrent une progression continue, avec \textbf{FCN} atteignant \textbf{66\%} et \textbf{68\%} respectivement. Cette amélioration s’accentue avec l’augmentation de la période d’observation, confirmant que l’enrichissement des données temporelles est crucial pour des prédictions précises, mais elle souligne aussi le défi du \textit{cold start} pour les semaines initiales. Les modèles classiques comme le \textbf{KNN} stagnent dans les deux configurations de classification, prouvant leur faible capacité à capturer des dynamiques temporelles complexes.
                             
Ainsi, que ce soit pour la classification binaire ou multiclasses, les cours présentant une richesse et une diversité élevées d’interactions temporelles offrent un avantage notable pour la prédiction des performances. À l’inverse, les cours avec une faible densité d’interactions posent des défis importants. La performance supérieure des modèles neuronaux, comme \textbf{FCN} et \textbf{DOPP}, illustre leur capacité à apprendre efficacement des séries temporelles multivariées, même si des méthodes adaptées restent nécessaires pour améliorer les performances dans des contextes où les données sont limitées ou faiblement structurées, comme l’utilisation de modèles préentraînés ou d’apprentissage par transfert.

Dans les cours \textbf{SHS}, des stratégies complémentaires telles que l’intégration de données qualitatives, comme les réponses des apprenants dans les forums de discussion ou les commentaires sur les évaluations, ou d’autres indicateurs comportementaux, tels que le temps passé sur les activités, la participation aux sessions interactives ou l’utilisation des ressources pédagogiques, pourraient améliorer les performances des modèles.

\begin{figure}[h]
    \centering
    \includegraphics[width=0.8\textwidth]{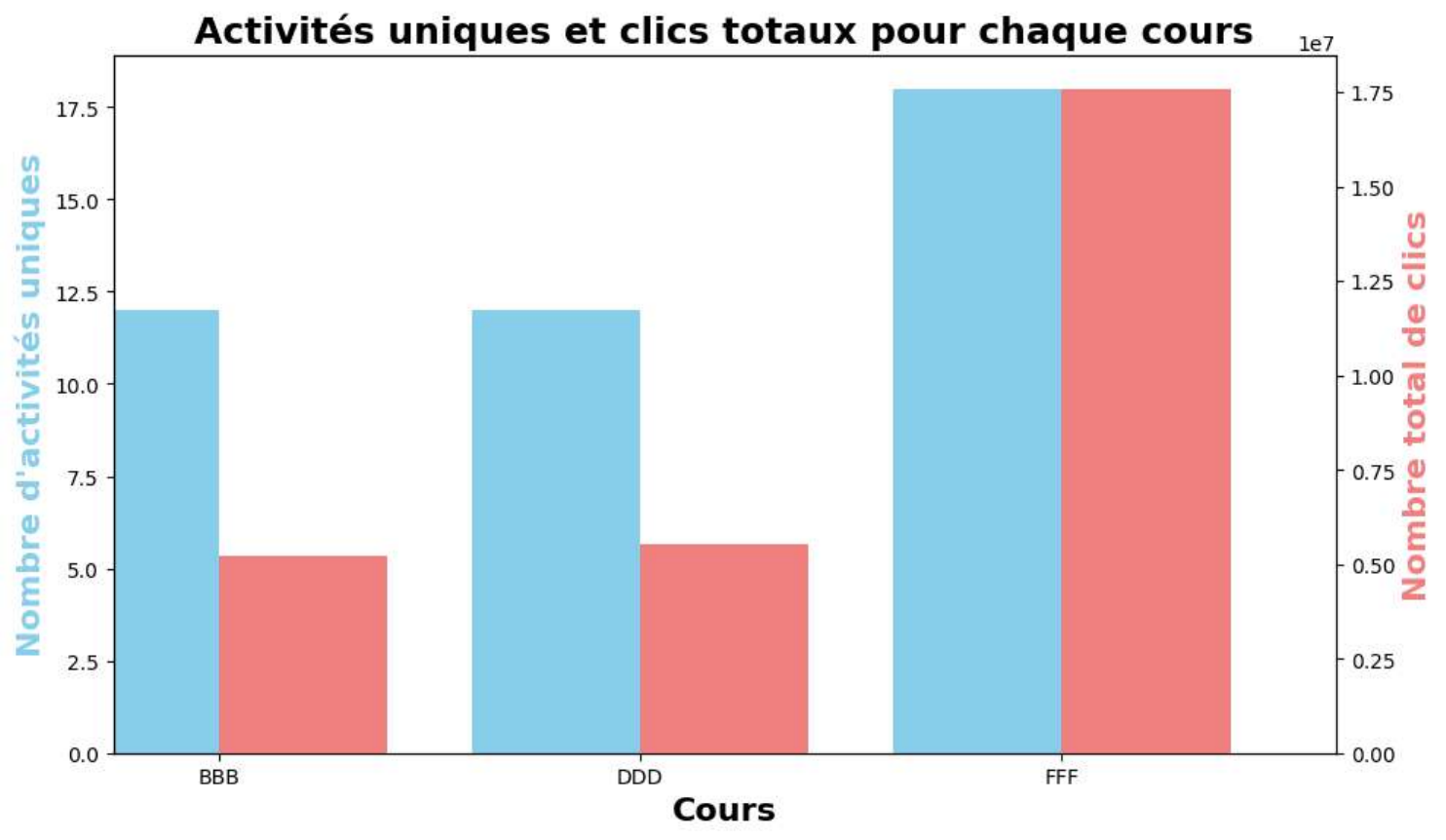}
    \caption{Comparaison du nombre unique d’activités et du total de clics pour les cours SHS et STEM. Cette figure met en évidence que le cours STEM dispose de davantage d'activités et d'interactions, suggérant une conception pédagogique plus riche et variée.
    }
    \label{fig:activity_clicks}
\end{figure}

La différence de performance observée entre les cours SHS et STEM peut être attribuée, en partie, à la conception pédagogique des cours. Une expérimentation complémentaire, illustrée dans (Fig.~\ref{fig:activity_clicks}), montre que le cours \textbf{FFF} (STEM) présente un nombre unique d’activités et un total de clics nettement supérieurs à ceux du cours \textbf{BBB} (SHS). Cette différence suggère que les cours STEM bénéficient d’une conception favorisant une plus grande richesse et diversité d’interactions. Toutefois, cette observation ne peut être généralisée sans des analyses plus approfondies pour mieux comprendre l’impact des différences structurelles dans la conception pédagogique sur les performances des apprenants.

\section{Conclusion}

En conclusion, cette étude met en évidence l’intérêt de l’analyse des séries temporelles multivariées pour classer les apprenants selon leur performance attendue dans les MOOCs. Les résultats montrent que la richesse et la diversité des interactions hebdomadaires jouent un rôle clé dans la précision des prédictions. Les modèles profonds, en particulier le FCN introduit ici, permettent de capturer des motifs temporels complexes, dépassant les approches classiques.

Ces travaux s’inscrivent dans la continuité des recherches en EIAH intégrant l’apprentissage automatique, avec une volonté d’innovation dans la personnalisation des parcours. À terme, des outils comme des systèmes d’analyse automatisée ou des tableaux de bord pour enseignants pourraient faciliter l’identification précoce des apprenants en difficulté et soutenir des interventions pédagogiques ciblées.

Cependant, cette étude présente certaines limites. Les conclusions sont basées sur un seul jeu de données (OULAD) et un nombre restreint de cours. Bien que ce dataset soit riche et structuré, des validations sur d’autres plateformes et contextes éducatifs sont nécessaires pour confirmer la généralisabilité des résultats.

Le problème du \textit{cold start}, marqué par une faible densité d’interactions durant les premières semaines, reste un défi. Nos essais avec des tailles de noyaux de FCN et la distance \textit{Dynamique Time Warping} (DTW) n’ont pas permis d’améliorations significatives. L’intégration d’informations contextuelles (telles que les profils démographiques et la date d’inscription) pourrait atténuer cet effet en apportant des indices supplémentaires lorsque les données comportementales sont insuffisantes.

Enfin, l’intégration de données qualitatives issues de forums, d’évaluations ou de retours libres représente une piste prometteuse pour enrichir les profils d’apprenants. Leur usage soulève cependant des défis méthodologiques liés à leur nature interprétative et déclarative. Des approches hybrides combinant données quantitatives et qualitatives, renforcées par des techniques comme le transfert d’apprentissage~\cite{tsiakmaki2020transfer}, pourraient améliorer la robustesse des analyses et leur adaptation à une diversité de contextes éducatifs.

\section*{Remerciements}

Cette étude s’inscrit dans le projet \textbf{COPCOT} (ANR-22-CE38-0003), soutenu par l’Agence Nationale de la Recherche (\textbf{ANR}), visant à personnaliser les contenus éducatifs numériques pour faciliter l’acquisition de compétences en informatique.
Nous remercions l’\textbf{ANR} pour son soutien financier et les partenaires du projet pour leurs contributions.

\label{sec:main_section}
\bibliographystyle{splncs04}
%\newpage
\bibliography{Bibliography}
%\break
%\printbibliography

\end{document}